\documentclass[12pt]{article}
\usepackage{times}
\usepackage{amsmath,amssymb}
\usepackage{epsfig,graphics,graphicx}
\usepackage{color,multicol}

\usepackage[normalem]{ulem}

\long\def\symbolfootnote[#1]#2{\begingroup
\def\thefootnote{\fnsymbol{footnote}}\footnote[#1]{#2}\endgroup}  

\usepackage[longnamesfirst]{natbib}
\usepackage{multirow}
\usepackage{color, colortbl}
\usepackage[margin=1in]{geometry}
\definecolor{Gray}{gray}{0.9}
\usepackage{setspace}

\usepackage[usenames,dvipsnames]{xcolor}

\begin{document}

\title{Categorical Data Fusion Using Auxiliary Information} 
\author{Bailey K. Fosdick, Maria DeYoreo, and Jerome P. Reiter}
\maketitle
\begin{abstract}

\renewcommand{\baselinestretch}{1.25}

\noindent
{In data fusion analysts seek to combine information from
  two 
  databases comprised of disjoint sets of individuals, in which some variables appear
  in both databases and other variables appear in only one database. 
  Most data fusion techniques rely on variants of conditional
  independence assumptions.  When inappropriate, 
   these assumptions can result in unreliable inferences. We propose a data
  fusion technique that allows analysts to easily incorporate auxiliary
  information on the dependence structure of variables not observed
  jointly; we refer to this auxiliary information as glue.  
 With this technique, we fuse two marketing surveys from the book publisher
 HarperCollins using glue from the online, rapid-response polling
 company CivicScience. The fused data enable estimation of 
 associations between people's preferences for authors and for learning
 about new books.  The analysis also serves as a case study on the potential for using online
 surveys to aid data fusion. 
}  
\vspace{2cm}
\end{abstract}


\symbolfootnote[0]{Address for correspondence: \texttt{bailey@stat.colostate.edu}, Department of Statistics, Colorado State University, 102 Statistics Building, Fort Collins, CO 80523-1877.  This research was 
supported by grants from the National Science Foundation (NSF-SES-11-31897 and DMS-1127914).  We thank members of the working group from the SAMSI program on Computational Methods in the Social Sciences, 
especially Nicole Dalzell, Elena Erosheva, Monika Hu, Tracy Schifeling, Joe Sedransk, and Aleksandra Slavkovic.  
We also thank David Boyle and Zach Sharek for providing the data from HarperCollins and CivicScience.}

\newpage

\renewcommand{\baselinestretch}{1.5}

\section{Introduction}
In many applications in marketing, analysts seek to combine information from
two or more databases containing information on disjoint sets of individuals and distinct sets of
variables \citep{Kamakura, putten2002, Kamakura2, gilula, van2008proof}.
For example, a company has one database on customers'
purchasing habits and another database on individuals' media viewing
habits, and seeks to find associations between viewing and
purchasing habits \citep{gilula}.  This 
setting, known as data fusion \citep[][p. 60 -- 63]{rasslerbook}, arises in other contexts, including microsimulation modeling in economics \citep{moriarty2003} and government statistics \citep{dorazio2002}. For applications in other areas, see \citet[][reprinted from a 1978
manuscript]{kadane}, \citet{rodgers:1984}, \citet{mor:fritz:2001}, and \citet{dorazio2006}.

Typical applications of data fusion rely on strong and unverifiable
assumptions about the relationships among the variables.  To see this,
consider fusion of two databases, $D_1$ and $D_2$, with disjoint sets of individuals.    
Let $A$ denote the set of variables common to both databases, such as
demographics; let $B$ denote the set of variables unique to $D_1$; and let $B'$ denote the set of variables unique to  $D_2$.  Since $\{A,B,B'\}$ are never observed
simultaneously, the joint distribution of $\{A,B,B'\}$ is not
identifiable based on $(D_1, D_2)$ alone. Neither is the distribution of
$\{B,B'\}$, either marginally or conditionally on $A$. 
Put another way, many possible specifications of the
joint distributions of $\{A,B,B'\}$ may be consistent with the
marginal distributions of $\{A,B\}$ in $D_1$ and $\{A,B'\}$ in $D_2$.
The data provide no information on which specifications to favor.

For data fusion to proceed, analysts must make some 
assumption about the joint distribution of $\{A,B,B'\}$.
The most common assumption is that the variables in $B$ are conditionally
 independent of those in $B'$, given the variables in $A$ \citep{kiesl, dorazio2006, gilula}. For example, assume that every person with the same
age, gender, occupation, race, county of residence, etc., has the same probability
of purchasing the product, regardless of their media viewing habits.
While this assumption could be reasonable in some contexts with rich
$A$ variables, it also could be grossly incorrect. 
  For example, in some demographic groups, people who watch advertising infrequently may be less
likely to purchase the product. When this is the case, assuming
conditional independence can result in  inferences about $\{A,B,B'\}$ that do not
 accurately reflect the underlying relationships in the population.

To reduce reliance on conditional independence assumptions, analysts
require some form of auxiliary information. For
example, analysts can use knowledge about the joint distribution of $\{B,B'\}$ from other sources 
to bound the joint distribution of $\{A,B,B'\}$ \citep{dorazio2006}. Another
possibility is to mount a new data collection that provides information
on unknown features of the
joint distribution of $\{A,B,B'\}$.  Historically, such surveys have been untimely and
prohibitively expensive.  However, in recent years technological
advances have opened the door to fielding rapid response, low cost
surveys  \citep{gilulamc}.  Questions then arise as to  how analysts can leverage 
the information in such surveys for more accurate data fusion.

In this article, we propose a data fusion approach that allows
analysts to incorporate auxiliary information on arbitrary subsets of
$\{A,B,B'\}$ with at least one variable in $B$ and $B'$ 
jointly observed.  We refer to such auxiliary information as 
\textit{glue}, since it serves to strengthen the connection between
$B$ and $B'$. We present the approach for the common setting of all
categorical variables, although similar strategies could be used for
numerical variables. 
The basic idea is to collect or construct a dataset that represents the auxiliary
information, append this dataset to the concatenated file $(D_1, D_2)$,
and fit an imputation model to predict missing $B$ in $D_2$ and
missing $B'$ in $D_1$.  As the engine for imputation, we use a Bayesian latent class model 
\citep{dunsonxing, si:reiter:13}.  
Using simulation studies, we illustrate how to 
accommodate glue of various sizes and on various variable subsets, and
demonstrate the potential for glue to improve
accuracy relative to fusion procedures that assume
conditional independence.  We also discuss problems that can arise when
using glue from a non-representative sample, and
propose methodology for incorporating non-representative glue
in data fusion. We
apply the methodology in a data fusion experiment in which we obtain
glue from the internet polling company CivicScience, and use the glue
to fuse surveys fielded by the book publisher HarperCollins Publishers on author preferences and author discovery tendencies.

 The remainder of the article is organized as follows.  In Section
 \ref{background}, we introduce the HarperCollins data fusion context
and review typical approaches to data
 fusion in the literature.  
In Section \ref{model}, we describe how to adapt Bayesian latent class models for data fusion to accommodate 
 glue. The approach allows for both the creation of completed data
 files, i.e., as in multiple imputation \citep{rubin:1986, rubin, reiter:12:sinica}, as well as
 parameter inference.  We focus on creating completed datasets, which
 can be subsequently analyzed using the techniques
 of \citet{rubin}. 
We also summarize results of simulation studies that demonstrate the benefits
of leveraging glue in data fusion.
In Section \ref{application}, we present results of the HarperCollins
Publishers' and CivicScience data fusion.
In Section \ref{conclusion}, we conclude with a discussion of open questions and future research directions.

\section{Background}\label{background}

\subsection{HarperCollins data and CivicScience glue}

HarperCollins Publishers routinely administers
surveys to the public to learn about their
behaviors and opinions, relying on this information to guide business decisions.  The surveys typically include 
questions about basic demographics (e.g., age, income, gender) and
reading habits, as well as questions on focused topics such as technology usage or author preferences.
Generally, around 10\% of questions in the surveys address basic demographics and reading habits, 
and the remaining 90\% are specific to the survey.  
We seek to fuse data from two HarperCollins surveys, one including
questions on the authors people read and the other including questions
on where people discover new authors (e.g., Facebook and Best Sellers lists). 
The first survey comprises $4,001$ respondents and $734$ variables;
we use only a subset of questions related to discovery and demographics.  
The second survey comprises $5,015$ respondents and $1,433$ variables; 
we use only a subset of questions related to author readership and demographics. 
The surveys were administered by
an independent company to a random sample of people residing in the United
States, with pre-specified 
numbers of individuals in specific categories based on age, gender,
ethnicity, and geographic regions.

HarperCollins is interested in understanding the demographics of readers of particular authors and how to reach them.  
For example, if HarperCollins publishes a new book by the author Lisa Kleypas,
will they reach more of her readers by advertising the new book in bookstores or on Facebook?  
Furthermore, who should be the target audience (age, gender, etc.) of the advertisements?  Leveraging the connections between 
author readership, book discovery, and demographics across surveys can help HarperCollins pursue profitable marketing strategies.

To obtain glue for the data fusion, we collaborated with internet polling company CivicScience.\footnote{Mark Cuban, the high-profile owner of the Dallas Mavericks and Shark Tank investor, was quoted in the
Pittsburgh Tribune Review in 2013 stating, ``CivicScience is one of
the most exciting companies I have seen in a long time. Their ability
to predict consumer behavior in media, retail sales, and even politics has virtually unlimited potential.'' }    
Internet polling companies are potentially ideal glue collectors, as they are able 
to survey thousands of people daily at low cost. 
As case in point, CivicScience collects hundreds of thousands of responses per day and has information stored on millions of respondents.
CivicScience is routinely paid by other companies to canvass the public on marketing and business decisions.  

CivicScience obtains information by posting short surveys, typically three or four questions, on the sidebar of
popular websites. 
Participation is purely voluntary (raising the potential for selection bias, which we return to later). 
CivicScience entices participation by beginning each survey with an engagement question that people are often willing and eager to share their opinion on (e.g., ``Who will win the Superbowl?").  The next question(s) is a value question asked on behalf of a paying client. 
 The final question inquires about 
respondent demographics.  After completing the short survey, participants are offered the option to answer additional questions.
CivicScience uses participants' computer IP addresses to link responses from the same individuals (more accurately, from the same computer).

For our application, CivicScience  
ran numerous three-question surveys on author readership and discovery. 
The second question was about either author readership or discovery, and the third question was about either the respondent's age or gender. 
Many participants completed more than one survey, allowing CivicScience to link responses on author readership, discovery, age, and gender.  We use these linked data 
in the fusion of the HarperCollins surveys.

\subsection{Common data fusion methods}

The most widely used data fusion technique in practice is statistical matching  \citep{putten2002,wicken2008}.  
The analyst divides the observations in $(D_1, D_2)$ into groups based on the similarity of values in the $A$ variables.  Within each group, the analyst
imputes missing $B$ values for records in $D_2$ by sampling from the empirical distribution of $B'$ in that group. The analyst imputes 
missing $B'$ values for records in $D_1$ in a similar manner. 
Often one cannot find groups of records in $D_1$ and $D_2$ with
exactly the same values on all of $A$, particularly when the contingency table
implied by the variables in $A$ has a large number of cells. 
In such cases, analysts form groups based on some subset of $A$
variables. Alternatively, analysts specify some distance function
 that quantifies how ``close" the $A$ values are for a given pair of observations from $D_1$ and $D_2$, 
  and form groups based on the close matches.  Regardless of how the
 analyst 
  forms groups, 
these approaches all make the unverifiable assumption that $B$ is independent of $B'$ within the analyst-specified groups.

A second approach to data fusion is to estimate regression models for the distributions of $(B \mid A)$ from $D_1$ and $(B' \mid A)$ from $D_2$, and 
set $f(B, B' \mid A) = f(B \mid A)f(B' \mid A)$, i.e., assume conditional independence between $B$ and $B'$ \citep{rodgers:1984,gilula}. 
One then imputes missing values of $B$ using the estimated model for $(B \mid A)$, and imputes missing values of $B'$ using the estimated model for $(B' \mid A)$.
\citet{gilula} describe how to adapt this regression-based approach to incorporate auxiliary information about the dependence between a single binary $B$ and a single binary $B'$.

A third approach is to estimate models for the entire joint distribution of $\{A,B,B'\}$. 
For example, one could use a multinomial distribution with probabilities constrained by a log-linear model that excludes terms 
involving interactions between $B$ and $B'$.  This also assumes conditional independence between $B$ and $B'$.  
\citet{dorazio2006} describe how this conditional independence assumption can be relaxed in log-linear models 
by incorporating auxiliary information on marginal probabilities for $(B, B')$.
Alternatively, one could estimate the joint distribution of
$\{A,B,B'\}$ with a latent class model \citep{goodman}, as suggested
by \citet{Kamakura} and as we do here.
Unlike log-linear models, latent class models can capture complex associations among the variables automatically, avoiding the   
 difficult task of deciding which interactions to include from the enormous space of possible 
models \citep{vermunt,si:reiter:13}.  Latent
class models also easily handle missing values in $D_1$ and $D_2$ due
to item nonresponse within the surveys, assuming nonresponse is missing at random \citep{rubin:1976}.
However, we are not aware of methodology for incorporating auxiliary
information when using latent class models in data fusion.
 We now introduce such methodology.

\section{Methodology}\label{model}
\subsection{Bayesian latent class models for categorical data fusion}
Suppose that we seek to fuse database $D_1$ comprising $n_1$
individuals with database $D_2$ comprising $n_2$
individuals. Let $Y_{ij}\in \{1,\dots,d_j\}$ be the value of variable $j$ for
individual $i$, where $j=1,\dots,p$ and $i=1, \dots, n_1+n_2$.  
Let $Y_i= (Y_{i1}, \dots, Y_{ip})$ for all $i$. 
The $p$ variables 
 form a contingency table with $\prod_{j=1}^pd_j$ cells. 
For variables $j \in A$, we observe $Y_{ij}$ for all $n=n_1 + n_2$
individuals; for variables $j \in B$, we observe $Y_{ij}$ for only the
$n_1$ individuals in $D_1$; and, for variables $j \in B'$, we observe $Y_{ij}$ for only the
$n_2$ individuals in $D_2$.  We note that, in practice, item
nonresponse will result in unintentionally missing values within $D_1$
and $D_2$ as well.

In latent class models for categorical data, we assume that each  
individual is a member of one of $N$ unobserved classes.  Let $Z_i \in \{1,
\dots, N\}$ denote individual $i$'s class membership, and let
$\pi_l=\mathrm{P}(Z_i=l)$ be the probability that individual $i$
is in class $l$. We assume that $\pi = (\pi_1, \dots, \pi_N)$  is the same for all
individuals.  Within each class, we assume the variables follow
independent categorical distributions with variable-specific probabilities
$\phi_l^{(j)}=(\phi_{l1}^{(j)},\dots,\phi_{ld_j}^{(j)})$, where 
$\phi_{ly}^{(j)} = \mathrm{P}(Y_{ij}=y \mid Z_i=l)$.  As a flexible and 
computationally convenient prior
distribution on $\pi$ and $\{\phi_l^{(j)}\}$, we use the truncated version of the Dirichlet
Process (DP) prior \citep{sethuraman}.
The complete model, referred to as the DP mixture of products of multinomials (DPMPM), can be expressed as: 
\begin{align}
Y_{i1},\dots,Y_{ip}|Z_i,\phi & \stackrel{ind.}{\sim} \prod_{j=1}^p
\mathrm{categorical}(Y_{ij};\phi_{z_i1}^{(j)},\dots,\phi_{z_id_j}^{(j)}),
\quad i=1,\dots,n \label{model:Y}\\ 
Z_i\mid \pi &\stackrel{ind.}{\sim} \mathrm{categorical}(\pi_1, \dots, \pi_N), \quad i=1,\dots,n \label{model:Z}\\ 
\pi_l & = V_l\prod_{r=1}^{l-1}(1-V_r),\,\,\, \pi_N=1-\sum_{l=1}^{N-1}\pi_l \nonumber \\
V_l\mid \alpha &\stackrel{iid}{\sim} \mathrm{beta}(1,\alpha), \,\, V_N = 1, \quad l=1,\dots,N-1  \nonumber \\
\phi_{l}^{(j)}&\stackrel{ind.}{\sim} \mathrm{Dir}(a_{1}^{(j)},\dots,a_{d_j}^{(j)}), \quad l=1,\dots,N, \: j=1,\dots, p  \nonumber \\ 
\alpha &\sim \mathrm{gamma}(a_{\alpha},b_{\alpha}) \label{model:alpha}.
\end{align}
The parameter $\alpha$
plays a central role in determining the number of effective components in the
mixture, with smaller values favoring fewer components. A hyperprior
on $\alpha$ allows the data to inform the number of components. In our
applications, we fix $a_\alpha$ and $b_\alpha$ equal 
to $0.5$ in the prior distribution in \eqref{model:alpha}, which
represents a relatively noninformative prior. We set
$a_1^{(j)}=\dots=a_{d_j}^{(j)}=1$ for all $j$.

We estimate the DPMPM model using Markov chain Monte Carlo (MCMC) posterior simulation
techniques 
\citep{ishwaranzar, ishjames}. 
The missing $Y_{ij}$,  unforeseen 
from item nonresponse and expected
due to the the structure of data fusion, are imputed as part of the MCMC. 
Given a draw of model parameters $(\alpha,
\{\phi^{(j)}\}, Z, V, \pi)$, we sample a value for each missing
$Y_{ij}$ from the relevant independent categorical distribution in
class $Z_i$. 
Further
details on the sampling algorithm are provided in the Appendix.

The probability model defined in 
\eqref{model:Y} and \eqref{model:Z} is the same as that 
 used by
\citet{Kamakura}.  However, rather than use a fully Bayesian estimation
  approach, they maximize the
  likelihood function obtained from equations (\ref{model:Y}) and
  (\ref{model:Z}).  Additionally, \citet{Kamakura} use heuristics to
  determine some optimal number of classes, whereas with the DPMPM one
  simply can fix the truncation level $N$
  to a large value \citep{ishjames}.
To ensure that $N$ is large enough, the analyst confirms that the number
of occupied classes $n^*$ is always significantly less than $N$ across
MCMC samples. If the posterior distribution for $n^*$ places
significant mass near $N$, then $N$ should be increased. In the
analyses in this article, $N=30$ is always sufficiently large.

Even though variables are independent within the latent classes, 
variables still can be marginally dependent across the set of classes. For
example, for any pair of variables $j$ and $j'$, we have 
\begin{equation}P(Y_{ij} = y, Y_{ij'} = y' \mid \pi,\{\phi^{(j)}\})=
  \sum_{l=1}^{N}\pi_l \phi_{l,y}^{(j)} \phi_{l,y'}^{(j')}. \label{eq:lc:joint}\end{equation} 
In general, the expression in \eqref{eq:lc:joint} is not identical to the product of
the two marginal probabilities,  
$\left(\sum_{l=1}^{N}\pi_l \phi_{l,y}^{(j)}\right)\left(\sum_{l=1}^{N}\pi_l \phi_{l,y'}^{(j')}\right)$, implying $Y_{ij}$ and $Y_{ij'}$ are independent conditional on $Z_i$ and $\{\phi^{(j)}\}$, but dependent upon marginalization over $Z_i$.
Expression \eqref{eq:lc:joint} can be used for model-based inferences
about probabilities.

As suggested by \citet{gilula} when discussing the model used by
\citet{Kamakura}, estimates of the joint distribution of  
$\{A, B, B'\}$ from latent class models may not be concordant with
conditional independence. In our simulations, we found that  
the DPMPM favors somewhat stronger correlation between $B$ and $B'$
than is implied under conditional independence.  This results 
from the clustering engendered by the DP prior specification, since the data contain no information
about $\{B, B'\}$ jointly.   
This finding underscores the potential benefits of using glue
when using latent class models for data fusion.

\subsection{Incorporating glue in data fusion}

\citet{schifeling:reiter} developed a strategy for incorporating prior
information about marginal probabilities into the DPMPM.  They suggest
constructing a hypothetical dataset that represents prior beliefs,
appending it to the collected data, and estimating the latent class
model with the concatenated real and hypothetical data.  As an
example, if one knows only that the true proportion of women in a population
is exactly 50\%, one can append a large hypothetical dataset with
equal numbers of men and women with all other variables missing. \citet{schifeling:reiter} show that
this approach fixes the posterior probability of being female at 50\%
without distorting the conditional distributions of other variables on
gender.  

We adapt this strategy to incorporate glue in data fusion.  We assume
that the analyst has glue data, $D_s$, in which some
subset of the $\{B,B'\}$ variables, possibly with $A$, is
measured.  For individuals $i= 1, \dots, n_s$ in $D_s$, let $Y_{i}$ be
the $p \times 1$ vector of measurements for the $i$th individual.  In
most data fusion scenarios, each $Y_i$ will be incomplete by design,
in that only some variables are available in $D_s$.  We assume that 
$Y_i$ for individuals in $D_s$ follows the model in \eqref{model:Y} --
\eqref{model:alpha}.  Thus, we concatenate $(D_1, D_2, D_s)$ in one
file, and estimate the DPMPM model using MCMC.   The information on
$\{A, B, B'\}$ available in $D_s$ influences the parameter
estimates, resulting in imputations of missing $B$ variables in $D_2$
and $B'$ variables in $D_1$ that reflect the dependence relationships
in the glue. For computational convenience, when fitting the MCMC we
impute missing values in $D_1$ and $D_2$, but not those in $D_s$.

The ideal glue includes data on all variables in $(A,B,B')$ and is a
sample from the distribution of $(A,B,B')$ in the population of
interest. In practice, glue may be available only on subsets of variables, such as
$(B,B')$. In addition, $D_s$ may not be representative of the
population. For example, in the HarperCollins and CivicScience
data fusion, only the conditional 
distributions $P(B\mid A, B')$ can be plausibly considered representative.

To investigate the potential benefits of glue in these scenarios, we
use three sets of simulation studies.  
First, we add glue on different subsets of variables to explore the
intuition that richer glue (i.e., glue that contains more variables
simultaneously observed) results in larger improvements in
inference.  Second, we analyze the sensitivity of inference to the
addition of varying amounts of data subjects in the glue.
Third, we study the validity of inferences when using glue
that is not representative of the population 
distribution of $(A,B,B')$.  We also present a method for appropriately
incorporating such information.  We note that each of these issues
arises when using the CivicScience data as glue.

\subsection{Simulation studies with representative glue}
\label{sec:simulations}
We simulate fusion settings using a third HarperCollins survey containing $4,000$ respondents and $1,056$ variables.
As the $A$ variables, we select demographics including gender, age, work status, and income.  As the $B$
and $B'$ variables, we select eBook reader ownership and number of hours spent reading per week, respectively.  
Table \ref{table:HCvariables} describes the variables in detail.  We create $D_1$ by
randomly selecting half of the $3,567$ complete cases and
removing reading hours, and create $D_2$ as the remaining
half of the complete case data with eBook reader ownership removed.  We are interested in fusing $D_1$ and $D_2$ to estimate the relationship between eBook reader ownership and reading hours per 
week, conditional on specific demographics variables.  Because we have the complete observations of $\{A, B, B'\}$  in the original data, 
we can compare results from data fusion to the ground truth.

To quantify the potential for glue in this example, we investigated the Fr\'{e}chet bounds \citep{dorazio2006} on $P(B=j,B'=k)$ for $j=1,2$ and $k=1,2,3$, as implied by the marginal distributions $P(A,B)$ and $P(A,B')$.  If these bounds are tight, signifying the probabilities are highly constrained by the observed marginal probabilities $P(A,B)$ and $P(A,B')$, then little is to be gained from incorporating glue.  Conversely, if the bounds on the cell probabilities of $P(B,B')$ are wide, glue has the potential to greatly improve inferences based on $P(B,B')$.  Note that the marginal distributions $P(B)$ and $P(B')$ themselves  constrain $P(B,B')$.  The Fr\'{e}chet bound widths on the six cell probabilities ranged from 0.163 to 0.169.  This implies that even with observing $\{A,B\}$ and $\{A,B'\}$ there remains a lot of uncertainty about $\{B,B'\}$, and potentially much to be gained from collecting glue.

\begin{table}[t]
\caption{Variables contained in the HarperCollins survey used for simulations.}
\centering
\begin{tabular}{ c || c | c }
 Variable & Group & Levels (Level Label) \\
  \hline                   
  \hline
  gender & $A$ & male (1), female (2)\\
  age & $A$ &  18-24 (1), 25-34 (2), 35-44 (3), 45-54 (4), 55-64 (5), 65+ (6) \\
  work status  & $A$ & emp FT (1), emp PT (2), homemaker (3), retired (4), self-employed (5), other (6) \\
  income & $A$ &  $<$25K (1), 25-45K (2), 45-75K (3), 75-99K (4), 100+K (5), won't say (6)\\ 
  eBook & $B$ &  yes (1), no (2)\\ 
  hours & $B'$ & $<1$ (1), 1-4 (2), 5+ (3)
\end{tabular}
\label{table:HCvariables}
\end{table}

\subsubsection{Glue richness} 
\label{sim:rich}

We consider four types of glue for $D_s$. In increasing order of richness, these include 
only the marginal distribution $\{B,B'\}$, the joint distribution of $\{A_g,B,B'\}$ 
where  $A_g$ represents gender, the joint distribution of $\{A_a,B,B'\}$ where $A_a$ represents age, and 
the joint distribution of $\{A_g,A_a,B,B'\}$.  In each case, we create glue by duplicating the appropriate variables 
for all respondents in the original survey; thus, $n_s = 3567$. 
We run the MCMC chains long enough to obtain $120,000$ posterior samples of all parameters.  From these runs, 
we sample $m=50$ completed datasets, $(D_1^*, D_2^*)$, which we use in
multiple imputation inferences.
 
To evaluate the impact of glue richness, we compare Hellinger
distances, which are commonly used to quantify the similarity between
two probability distributions \citep{pollard, gibbs}.  Hellinger
distances based on $\{A,B,B'\}$ reflect the accuracy of the entire estimated joint distribution
$P(A,B,B')$, which arguably is the most important level of validity a
fusion process can achieve \citep{rassler}.
For two discrete distributions $P$ and $Q$ taking on $k$ values with probabilities $(p_1,\dots,p_k)$ and 
$(q_1,\dots,q_k)$, the Hellinger distance is given by $2^{-1/2}\sqrt{\sum_{i=1}^k (\sqrt{p_i}-\sqrt{q_i})^2}$.
 This quantity is between zero and one, where smaller values imply more similarity between the distributions.  Because the 
richest type of glue contains observations on $\{A_g,A_a,B,B'\}$, we compute Hellinger distances between the 
empirical distribution of $(A_g,A_a,B,B')$ based on the original complete survey and the corresponding posterior inferences.  
Calculations of distances based on the joint distribution $(A,B,B')$ including all demographic variables, rather than just $(A_g,A_a,B,B')$, yield similar patterns.

Table \ref{table:hell_dist} displays the posterior means and $95\%$ credible intervals for the Hellinger distances
 between the empirical distribution of $(A_g,A_a,B,B')$ and the corresponding posterior estimates. The results indicate that
   using glue can yield significant gains in accuracy, with increasing
   gains with richer glue. These results also suggest that gender
   offers smaller gains than age, a consequence of the fact that the
   distribution of $\{B,B'\}$ is more similar across gender than
   age. This finding is evident in all of the evaluations that
   follow.  Table \ref{table:hell_dist} also displays results from a set
   of fused data files using an exact matching algorithm
   based on all variables in $A$. The empirical joint
 probability distribution is comparable to that produced from the
 latent class model with no glue.

We also compare the sum of the absolute differences between the counts in the true contingency table for $\{A_g,A_a,B,B'\}$ based 
on the original complete data file and those based on imputed complete
data files.    These counts, when divided by two, indicate how many individuals the model places in incorrect cells of the empirical contingency table. We approximate the expected number of ``misclassified" individuals in an imputed data set with the empirical average over $50$ imputed data files.  Mathematically, the approximation for the expected number of misclassified individuals can be expressed
$$\mathrm{E}\left(0.5\sum_{j=1}^{\prod_{k=1}^pd_k}|n_{j}-\hat{n}_{j}| \right) \approx \frac{1}{50} \sum_{m=1}^{50} \left( 0.5\sum_{j=1}^{\prod_{k=1}^pd_k}|n_{j}-\hat{n}^{(m)}_{j}| \right),$$
where $\hat{n}^{(m)}_{j}$ is the number of individuals in cell $j$ in the $m$th imputed data set and $n_{j}$ is the true number of individuals in the original complete data set.  
Table \ref{table:incorrect_cell_counts} shows similar patterns as Table \ref{table:hell_dist}: using glue improves over
approaches that assume conditional independence, with increasing gains as the glue becomes richer.
We note that adding gender information to glue already containing age
does not lead to much improvement in imputation accuracy.

As a more focused evaluation, we use the completed datasets to estimate a logistic regression of eBook reader ownership on reading hours 
 and the demographics variables.  
The model includes terms for all main effects for all predictors, pairwise interactions between reading hours and gender and reading hours and age, and 
the three way interaction among reading hours, gender, and age.
Letting $A_i$ represent income and $A_w$ represent work status, the link function can be expressed as 
\begin{align*}
\mathrm{logit}(p(&B=1))= \beta_0+\beta^g1(A_g=2)+\sum_{k=2}^{6}\beta_k^a1(A_a=k)+\sum_{k=2}^6\beta_k^w1(A_w=k)\\
&+\sum_{k=2}^{6}\beta_k^i1(A_i=k)+\sum_{k=2}^3\beta_k^h1(B'=k)+\beta^{gh}1(A_g=2,B'=3)\\
&+\beta^{ah}1(A_a=6,B'=3)+\beta^{gah}1(A_g=2,A_a=6,B'=3).
\end{align*}

\begin{figure}

\centering
\includegraphics[width=.45\textwidth,height=.42\textwidth]{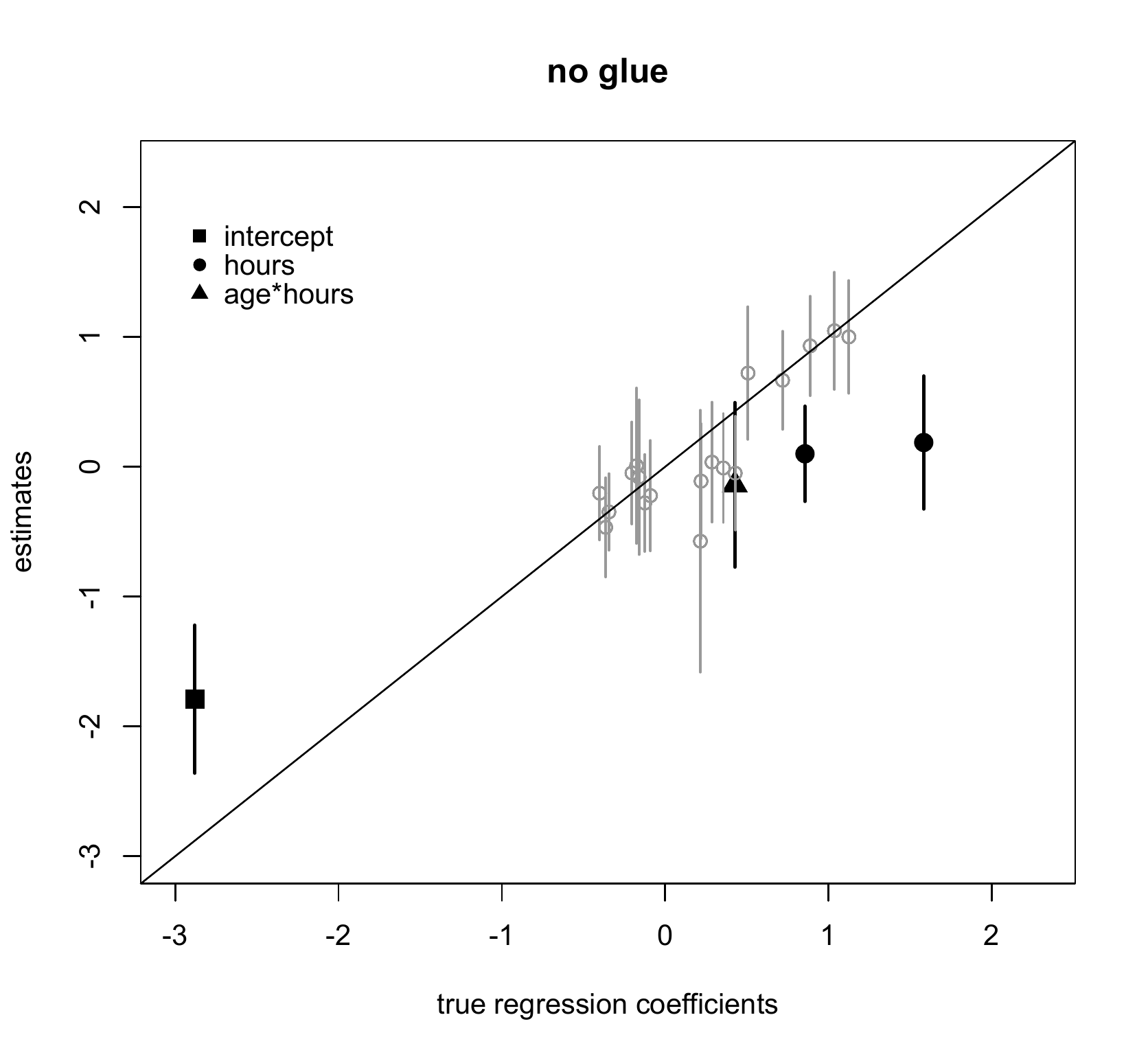}\\
\begin{tabular}{cc}
\includegraphics[width=.45\textwidth,height=.42\textwidth]{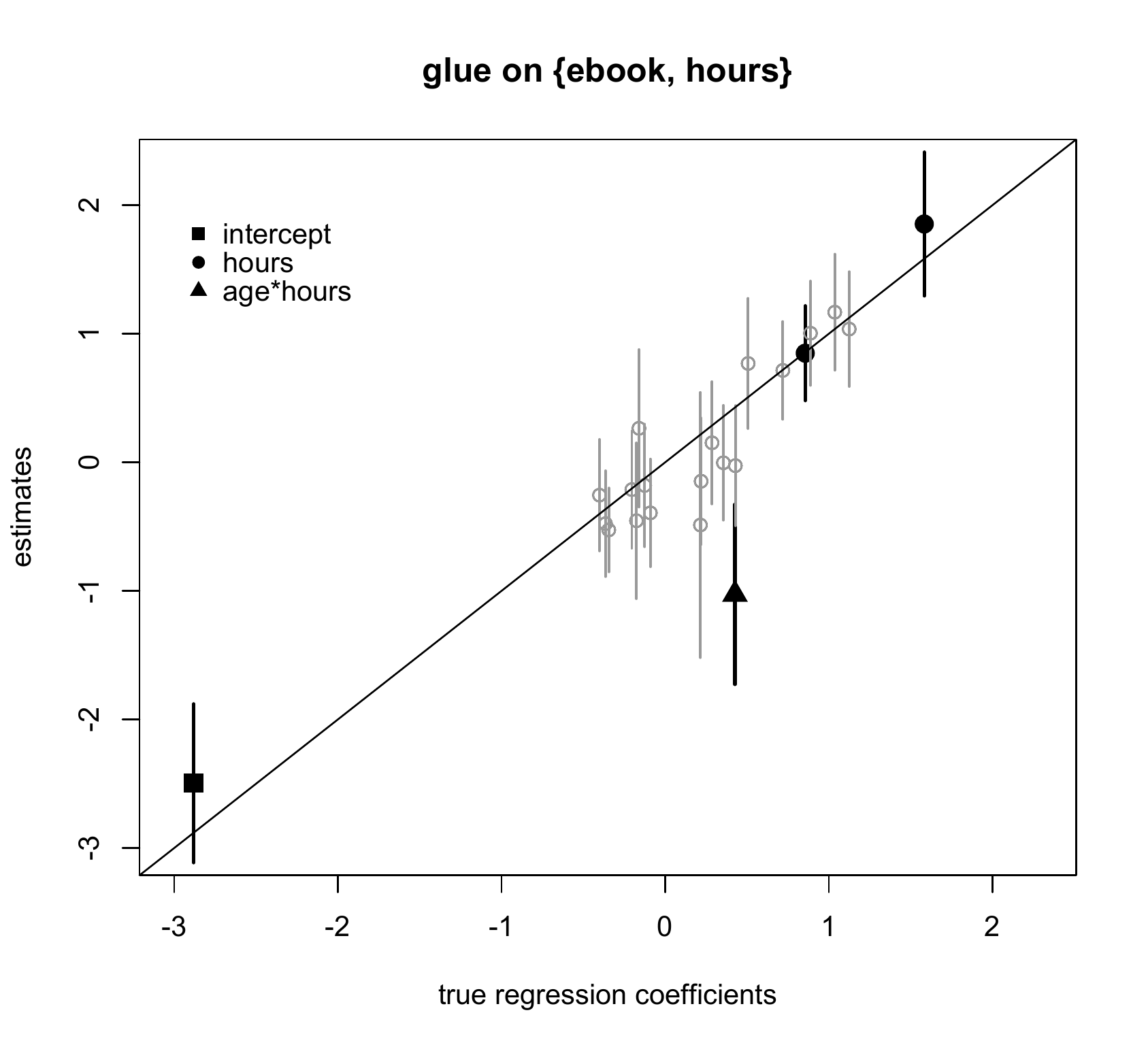}&
\includegraphics[width=.45\textwidth,height=.42\textwidth]{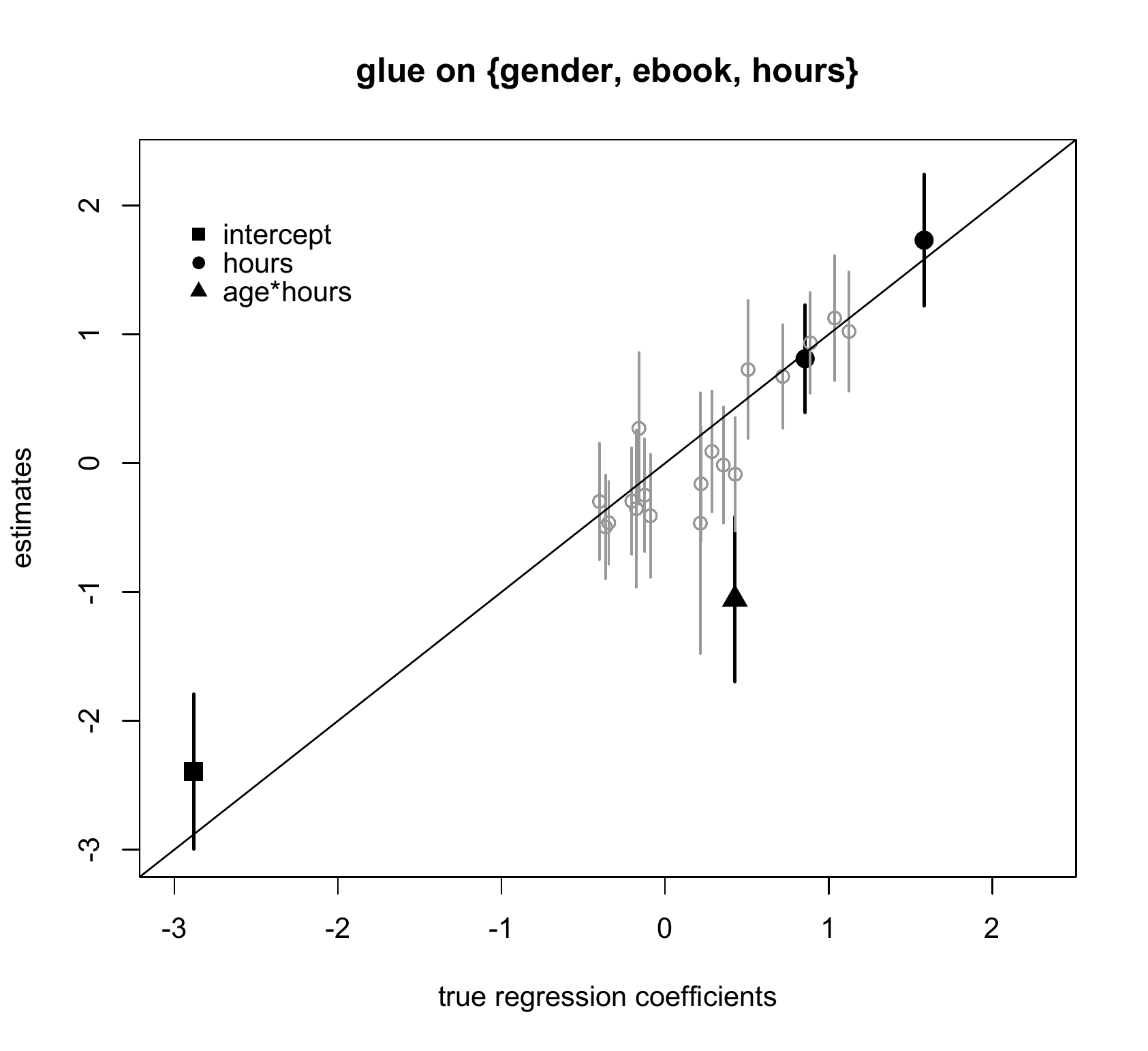}\\
\includegraphics[width=.45\textwidth,height=.42\textwidth]{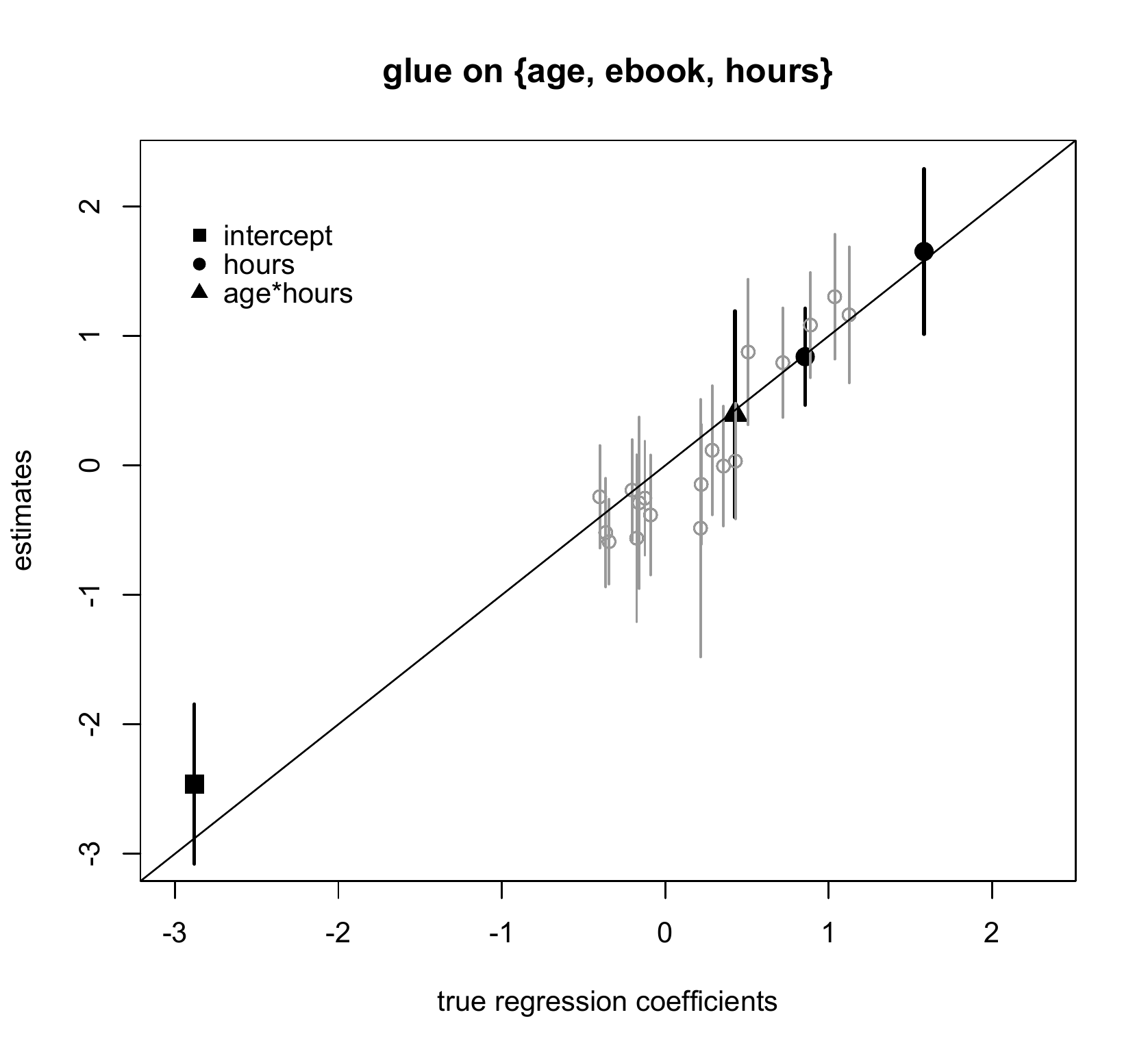}&
\includegraphics[width=.45\textwidth,height=.42\textwidth]{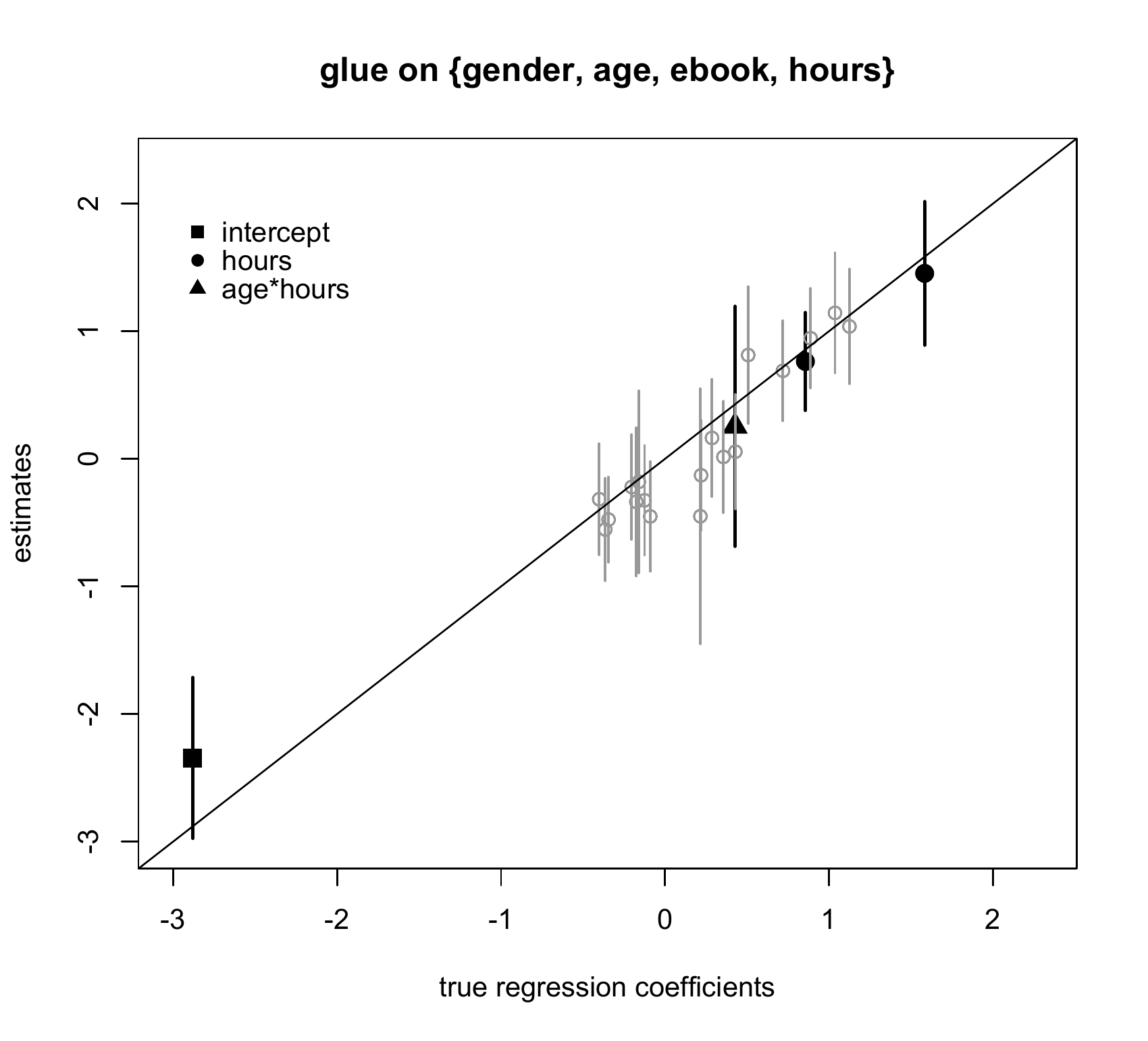}
\end{tabular}
\caption{Point estimates and $95\%$ confidence intervals for estimated versus true regression coefficients under five different glue scenarios. The first plot refers to the no glue scenario, and highlights terms which are affected by adding glue. These same $4$ terms are highlighted in the remaining plots as more glue is added.}
\label{fig:reg_coefs}
\end{figure}

We estimate the coefficients from the $50$ completed data sets using the standard 
multiple imputation combining rules \citep{rubin}.  As displayed in Figure \ref{fig:reg_coefs}, 18 of the 
22 regression coefficients based on the original data are contained in the 95\% MI confidence intervals under the data fusion model applied with no glue.
All intervals contain the original data coefficients when glue includes $\{A_a,B,B'\}$ as well as $\{A_g,A_a,B,B'\}$. 
Adding glue with only $\{B,B'\}$ improves the estimates of the main effects associated 
with $B'$ (reading hours).  Adding glue with at least $\{A_a,B,B'\}$ results in further improvements, in particular resulting in  
more reliable estimates of the interaction term associated 
with $A_a \times B'$ (age $\times$ hours). Clearly, even targeted
inferences can be improved by collecting glue, with generally
increasing gains with richer glue.

\subsubsection{Glue size} 
\label{gluesize}

In Section \ref{sim:rich}, the glue sample size was equal to the total
survey sample size, that is, $n_s = n = 3567$. 
Generally, this will not be the case.  To evaluate the role of glue sample
size, we repeated the simulations using $\{A_g, A_a, B, B'\}$ as glue
with different sample sizes for $D_s$. 
As shown in Table \ref{table:B_glue_sizes}, as expected, more high quality glue observations result in more accurate estimates with less uncertainty. 
Data fusion with $n_s = 1784$ glue cases yields inferences that are
close to the ground truth and to the inferences produced with more glue cases, suggesting 
that even modest amounts of glue can improve inferences.

\begin{table}
\caption{Posterior mean and $95\%$ credible intervals for the Hellinger distance between the true and estimated probability table for $(A_g,A_a,B,B')$ under five different glue scenarios, as well as the estimate obtained from a fused data set under statistical matching. *The range of Hellinger distances across 10 perfect matchings is reported to quantify matching uncertainty. }
\centering
\begin{tabular}{ c || c | c }
 & mean & $95\%$ CI or range* \\
  \hline                   
  \hline
   No glue & .104 & (.094,.113) \\
  $\{B,B'\}$ & .083 & (.075,.091) \\ 
  $\{B,B',A_g\}$ & .077 &  (.071,.084)\\ 
  $\{B,B',A_a\}$ & .060 & (.053,.068) \\ 
  $\{B,B',A_g,A_a\}$ & .052 & (.047,.059) \\
Exact matching & .100 & .090 - .107 \\
\end{tabular}

\label{table:hell_dist}
\end{table}

\begin{table}
\caption{Average number of individuals in the incorrect cell of the contingency table across the complete data sets under five different glue scenarios and under statistical matching.  Ten complete data sets were considered for the statistical matching procedure.}

\centering
\begin{tabular}{ c || c  }
 &  $\mathrm{E}\left(0.5\sum_{j=1}^{\prod_{j=1}^pd_j}|n_{j}-\hat{n}_{j}| \right)$ \\
  \hline                   
  \hline 
  no glue &  318.5 \\
  $\{B,B'\}$ &  250.5 \\ 
  $\{B,B',A_g\}$ & 247.0   \\ 
  $\{B,B',A_a\}$ &   199.5 \\ 
  $\{B,B',A_g,A_a\}$ & 196.0\\
  Exact matching & 315.0\\
\end{tabular}
\label{table:incorrect_cell_counts}
\end{table}

\begin{table}
\caption{Posterior mean and width of $95\%$ credible intervals for the marginal bivariate distribution of $P(B,B')$ under three different glue sample sizes.}
\centering
\begin{tabular}{ c || c | c | c | c}
 & truth & $n_s=0$ & $n_s=1784$ & $n_s=7135$ \\
  \hline                   
  \hline
  $P(B=1,B'=1)$ & .037 & .077 (.021) & .042 (.018) &  .040 (.009) \\
 $P(B=2,B'=1)$ & .363 & .333 (.041) & .357 (.033) &   .362 (.020)\\ 
  $P(B=1,B'=2)$ & .064 & .067 (.016) &.072 (.019) &   .066 (.011)\\
  $P(B=2,B'=2)$ & .252 & .248 (.036) & .247 (.030) &   .251 (.019) \\ 
  $P(B=1,B'=3)$ & .096 &  .062 (.017)& .089 (.020) &   .093 (.012) \\
  $P(B=2,B'=3)$ & .186 & .213 (.036) &.192 (.027) &   .188 (.017) \\
\end{tabular}

\label{table:B_glue_sizes}
\end{table}

\subsubsection{Nonrepresentative glue} 
\label{nonrepglue}

While glue obtained from non-probability samples like CivicScience
polls is convenient and inexpensive, it generally is not representative of the joint distribution of $\{A, B, B'\}$
in the target population for $(D_1, D_2)$. 
For example, $D_s$ may disproportionately represent some demographic groups compared to their shares in $(D_1, D_2)$. 
When the concatenated data $(D_1, D_2, D_s)$ is not a (incomplete) draw from 
$P(A, B, B')$, the posterior distributions of the DPMPM model
parameters will not produce accurate estimates of $P(A, B, B')$.    
The resulting imputations will be draws from a biased estimate of $P(A, B, B')$, which can diminish or even negate the benefits of using glue.  In various simulations, not reported here to save space, 
we found that significant problems can arise when appending nonrepresentative glue, even when the glue is representative 
of the population in terms of $P(B,B'|A)$ but not representative in terms of $A$.

When $D_s$ is not representative of the population, one still can construct useful glue provided that either $P(B \mid B', A)$ or
$P(B' \mid B, A)$ in $D_s$ is a draw from the corresponding conditional distribution in the population. The analysis proceeds as follows. 
\begin{enumerate}
\item
Fit the DPMPM model to $D_s$ alone to estimate $P(A,B,B')$, from which one can obtain $P(B|A,B')$ and $P(B'|A,B)$. 
\item
Construct glue $D_s^*$ by duplicating or sampling records $\{A,B\}$ with replacement from $D_1$,
or duplicating or sampling records $\{A,B'\}$ with replacement from $D_2$, and imputing the
missing values of $B'$ from $\{B' |A,B\}$ 
and the missing values of $B$ from $\{B|A,B'\}$ based on the conditional distributions from step (1). 
\end{enumerate} 
In this way, the constructed glue appropriately reflects the marginal distribution of $A$ and the information in the conditional distributions. 
With glue representing the appropriate joint distribution, we are in the scenarios described in Section \ref{sim:rich} and 
Section \ref{gluesize}.

To assess the validity of the assumptions that $P(B|A,B')$ and $P(B'|A,B)$ from $D_s$ are representative of the population of interest,
analysts can compare the empirical distributions of the sampled $B$ and $B'$ variables in step (2) to those from $D_1$ and $D_2$. 
When these empirical distributions differ greatly, the assumptions of conditional representativeness of the glue may be inappropriate, and 
the glue is not useful for data fusion. 
 When only one conditional distribution, either $P(B|A,B')$ or $P(B'|A,B)$, seems reasonable, the glue can be constructed 
using that conditional distribution only. Analysts can choose the number of records in the constructed $D_s^*$ to reflect their level of 
certainty about the conditional distributions. As a default, we recommend using the same sample size as the collected $D_s$.

We now illustrate that this diagnostic procedure can detect whether or not glue is representative on $P(B \mid A, B')$ or 
$P(B' \mid A, B)$. We consider a setting in which $D_s$ is representative on $P(B \mid A, B')$ but 
not on $P(B' \mid A, B)$, constructed as follows.  For $\{A_g,A_a\}$,
 we over-sample women and older individuals by keeping all 
 observations with $A_g=2$ or $A_a>4$, and sample each of the remaining observations with probability $0.5$.  This results in 
 $n_s=2,837$ auxiliary cases. We sample each record's $B'$ from $\{1,2,3\}$ with probabilities $(0.7,0.15,0.15)$.  This is 
highly nonrepresentative, as the true marginal probabilities are $(0.41,0.32,0.27)$. We sample each record's $B$ from $\{1,2\}$ 
with probabilities given by the empirical $P(B\mid A,B')$ from the original data. Thus, $D_s$ is representative in terms of
 $P(B\mid A,B')$, but not on $P(B'\mid A,B)$ or any marginal distributions.
We fit the DPMPM model to $D_s$ to estimate $P(B\mid A,B')$ and $P(B'\mid A,B)$, as described 
in step $(1)$, and construct $D_s^*$ as described in step $(2)$. The resulting marginal distribution for the 
imputed $B$ is extremely close to the empirical distribution of $B$ from $D_1$, with differences of only $0.01$. The marginal 
distribution for imputed $B'$ is $(0.57, 0.23, 0.20)$, quite far from the original data values. The diagnostic 
suggests that $P(B'\mid A,B)$ is not representative, whereas it may be reasonable to rely on $P(B\mid A,B')$.

\section{HarperCollins data fusion with CivicScience glue}\label{application}
We now turn to the HarperCollins data fusion.  We seek to combine
information from two surveys. In $D_1$, HarperCollins asked $n_1=2,000$ respondents questions related to the
discovery of new authors, e.g., ``Do you become aware of an author by
[medium]?'' for different mediums.\footnote{Although the survey contained $4,001$ respondents, only half were asked about author discovery.} In $D_2$,  HarperCollins asked
$n_2=5,015$ different people about their interest in various authors.
Each person was asked about different subsets of authors, so $D_2$
includes many missing values. We let $B$ represent author discovery via the
mediums Best Seller List, Facebook, library, online, recommendations,
and bookstore.  We let $B'$ represent interest in the 
authors Shel Silverstein, Agatha Christie, Suzanne Collins, Stephenie
Meyer, and Lisa Kleypas.  Each $B_j$ is recorded as yes or no. 
Each $B_j'$ is recorded as one of three categories, namely read,
interested, or not interested. Both $D_1$ and $D_2$ 
contain the demographic variables age, gender, and income, all of
which are of strong interest  to HarperCollins for market
segmentation. Our goal is inference on relationships between discovery
medium and author interest, in particular on the distributions $P(B|B')$, $P(B,B')$, and
$P(B,B'|A)$.

\begin{figure}
\centering
\includegraphics[width=5in,height=3.5in]{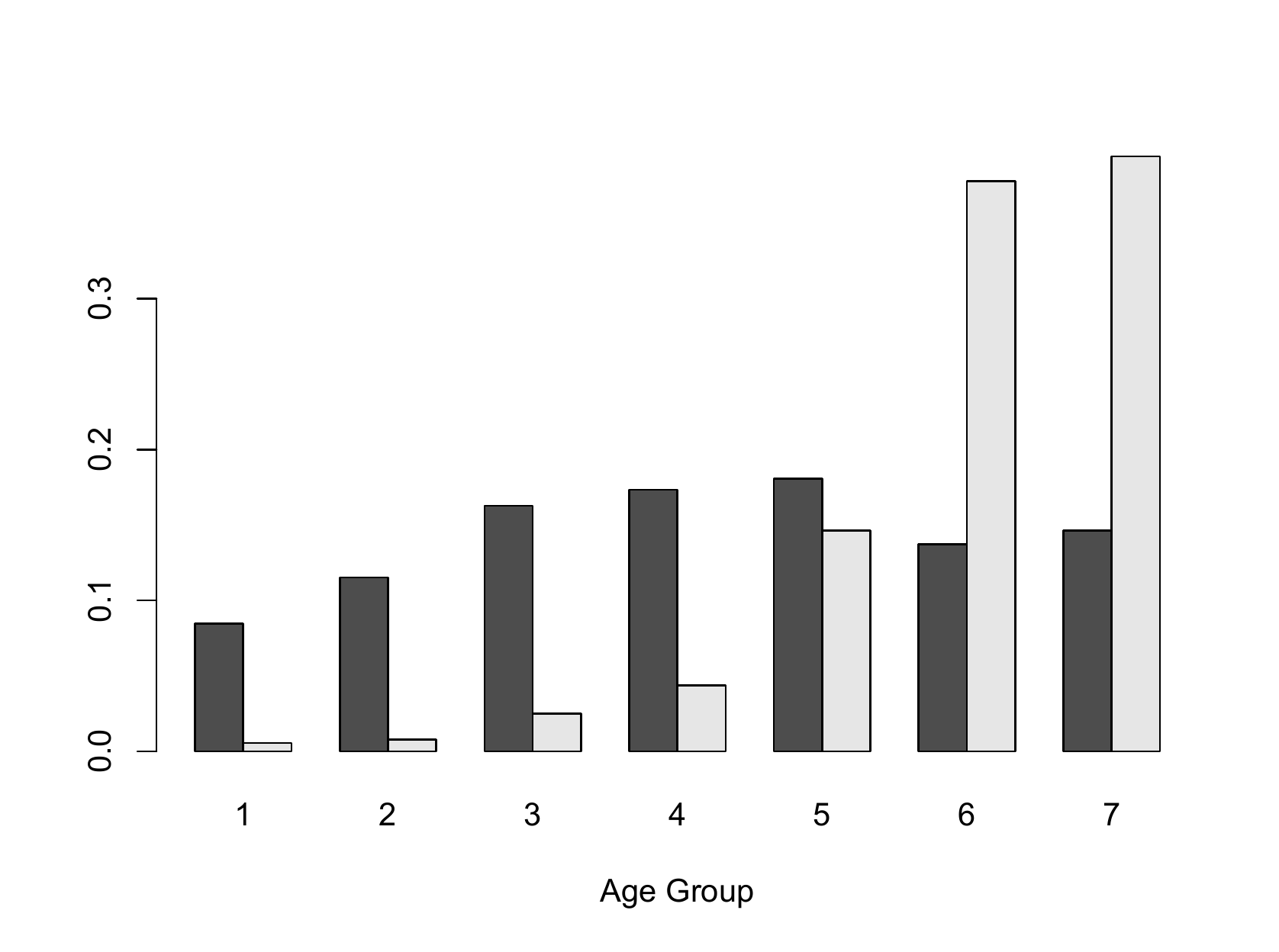}
\caption{Age distributions from the HarperCollins (dark gray) and CivicScience (light gray) surveys. }
\label{figure:age_dist}
\end{figure}

We provided CivicScience with a list of questions to ask in one of their 
surveys, with the goal of procuring glue. CivicScience 
collected $n_s=2,730$ simultaneous observations on author discovery and
interest, along with age and gender for many (but not all)
respondents.  
There are some key differences between the 
data collected by CivicScience and those in the original HarperCollins
surveys. In particular, the CivicScience respondents tend to be 
older; over $60\%$ are $55+$ years old compared to only $30\%$ of
HarperCollins respondents (see Figure \ref{figure:age_dist}). We
conjecture that is a consequence of the voluntary nature of the
internet data collection done by CivicScience.  
We note that the distributions of $A$ variables in $D_1$ and $D_2$ are very similar.

\begin{figure}
\centering
\includegraphics[width=3in,height=3in]{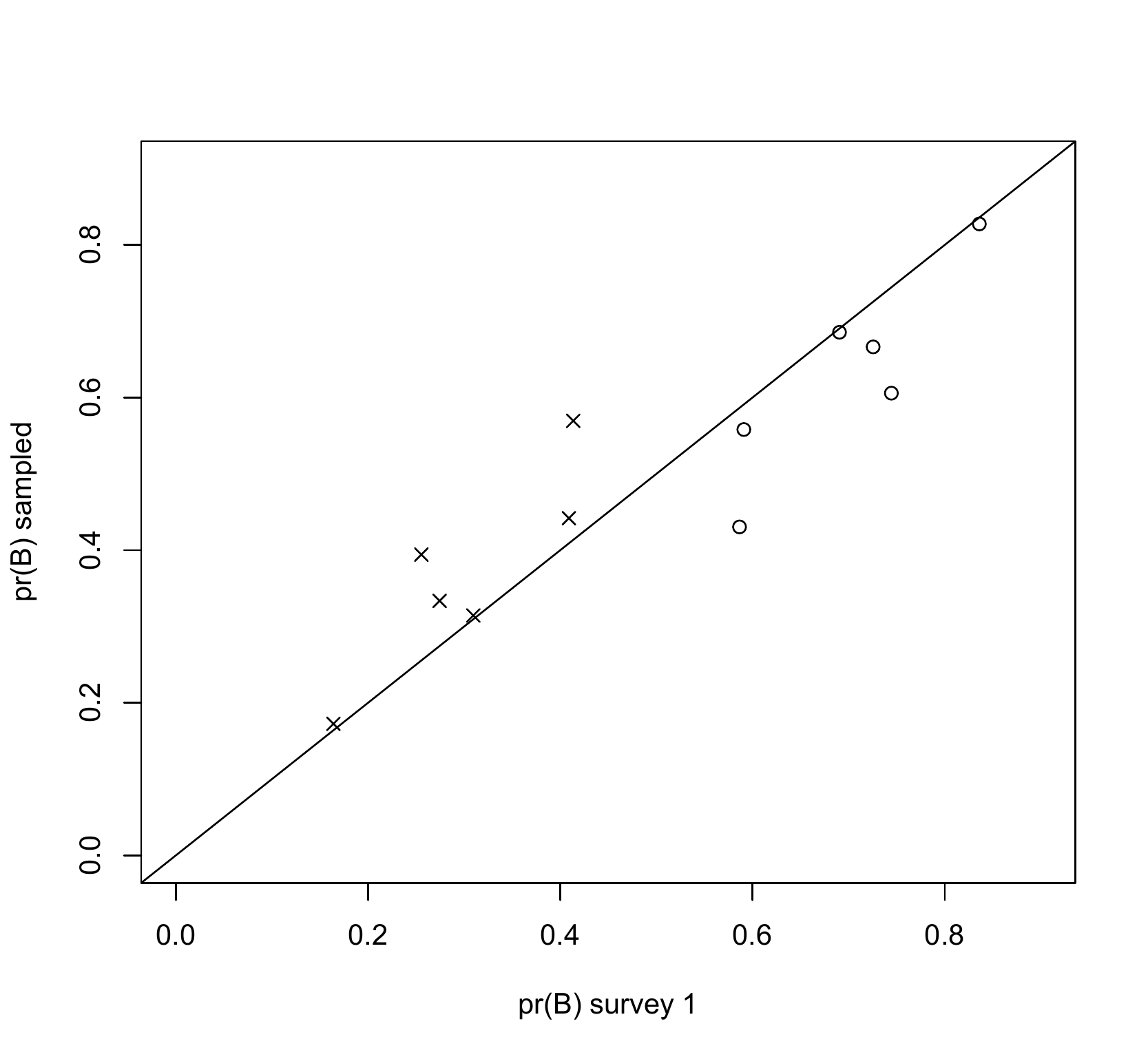}
\includegraphics[width=3in,height=3in]{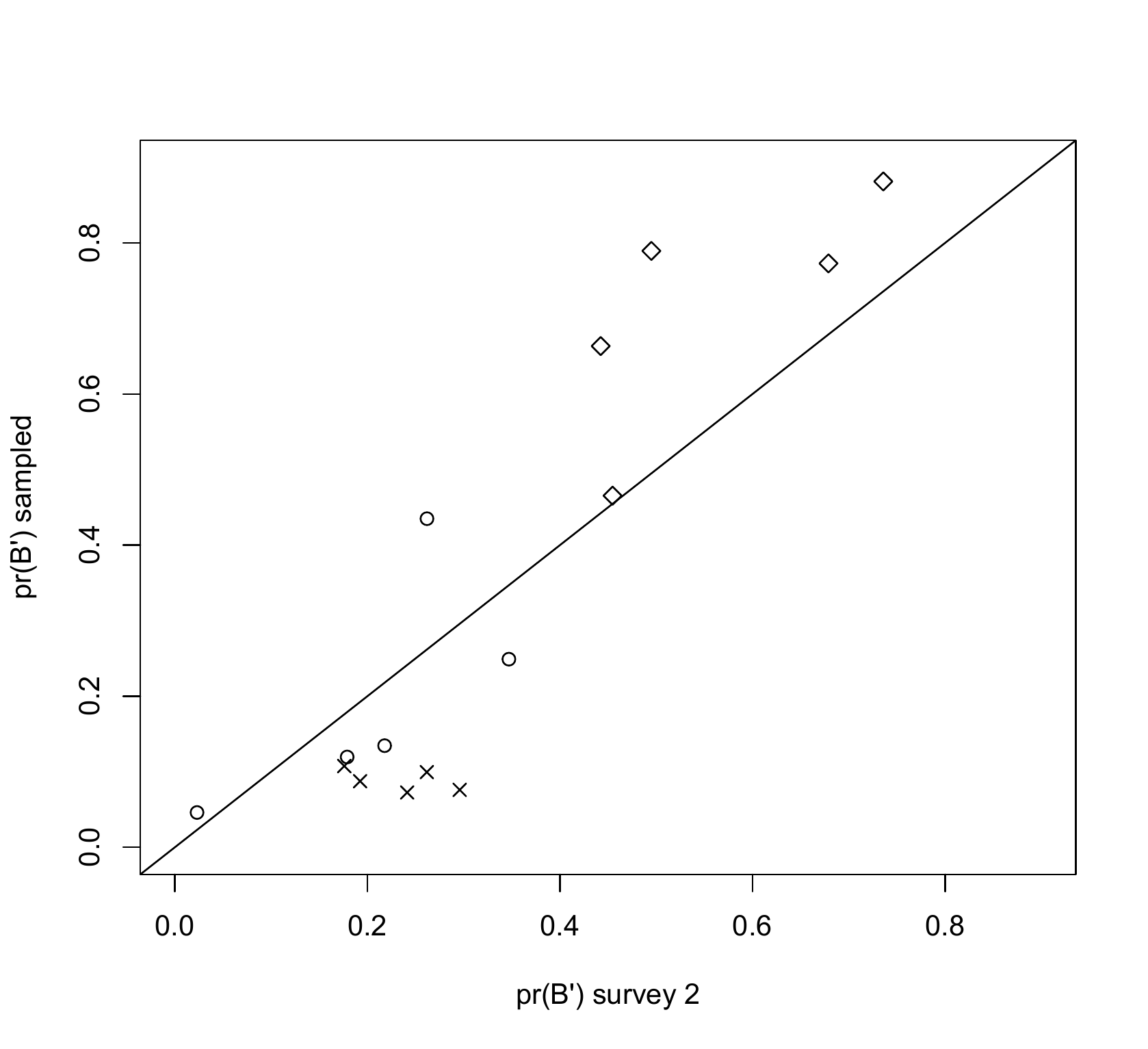}

\caption{Left: Empirical probabilities assigned to category 1 (`o' symbol) and category 2 (`$\times$' symbol) for each of 6 discovery questions by sampling $B$ as implied by inference for $P(B|A,B')$ from the CivicScience data versus marginal distributions of $B$ from the survey data. Right: Empirical probabilities assigned to category 1 (`o' symbol), category 2 (`$\times$' symbol), and category 3 (`$\diamond$' symbol) for each of 6 author interest questions by sampling $B'$ as implied by inference for $P(B'|A,B)$ from the CivicScience data versus $B'$ from the survey data.}
\label{figure:B_Bprime_sampled}
\end{figure}

As discussed in Section \ref{sec:simulations},  it is not prudent to
proceed with data fusion by appending the non-representative sample from the CivicScience survey to $(D_1, D_2)$. 
We therefore construct $D^*_s$ that reflects the marginal distribution of $\{A, B'\}$ in $D_2$ and the 
conditional distribution $P(B \mid A, B')$ estimated from the collected CivicScience data, following 
the procedure for non-representative glue described in Section \ref{nonrepglue}.  We first duplicate $\{A,B'\}$ from $D_2$,
and then sample values of $\{B|A,B'\}$ for these duplicated records using a DPMPM
applied to the CivicScience data. 
As evident in Figure \ref{figure:B_Bprime_sampled}, the empirical probability
distributions for the observed values of $B$ in $D_1$ and the sampled values of $B$ from 
$P(B\mid A, B')$ are  similar, suggesting that it is not unreasonable to use the CivicScience 
data to estimate $P(B\mid A, B')$.
We also considered creating $D^*_s$ by duplicating $\{A, B\}$ from $D_1$ and sampling $\{B'|A,B\}$ for the
duplicated records.  However, as shown in Figure \ref{figure:B_Bprime_sampled}, the sampled marginal distributions for
$B'$ do not closely match the empirical distributions in $D_2$.
We therefore do not assume $\{B'|A,B\}$ in the CivicScience data is
representative, and construct $D^*_s$ only from the duplicated $\{A,B'\}$ sample from $D_2$.

After appending the constructed $D^*_s$ to $(D_1, D_2)$, we estimate the 
DPMPM model on the concatenated data.  In the process we impute all missing values in $D_1$ and $D_2$.  As in the simulation studies, we 
keep $m =50$ of these completed datasets, spacing them far apart in the MCMC iterations to ensure approximate independence.  We use the completed versions 
of $D_1$ and $D_2$ for multiple imputation inferences.

\begin{figure}
\centering

\includegraphics[width=6.5in,height=3.3in]{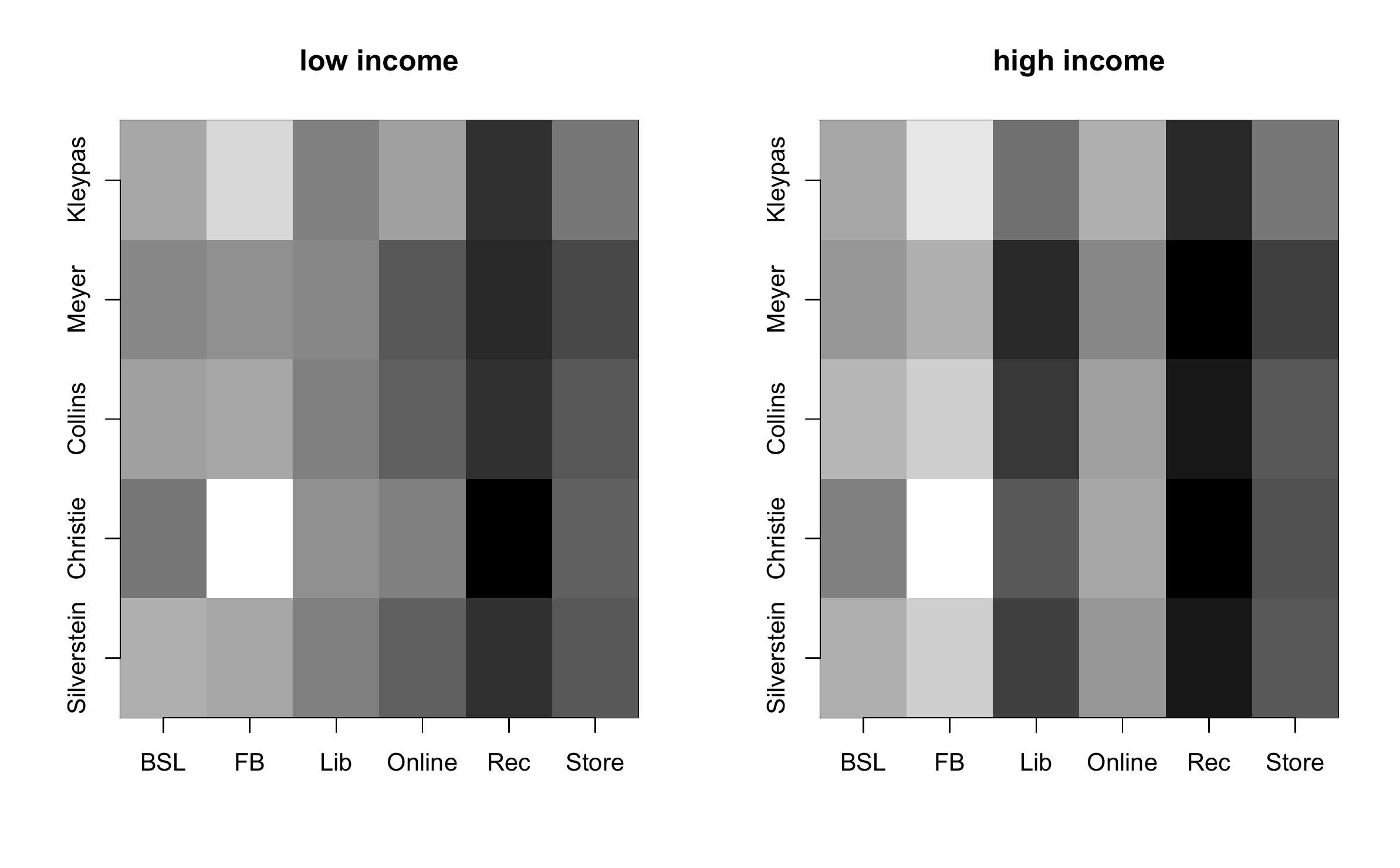}

\caption{Multiple imputation point estimates for $P(B=\mathrm{yes}|B'=\mathrm{read},\mathrm{income})$ for low and high income groups and all mediums $B$ and authors $B'$. Black indicates larger probabilities, and white indicates smaller probabilities.}
\label{figure:B_Bprime_cond_A_image}
\end{figure}

As a first data fusion inference relevant for marketing strategies, we estimate probabilities of discovery
via a given medium for those who have read or are interested in
reading a particular author. 
As evident in Figure \ref{figure:B_Bprime_cond_A_image},
high income individuals appear very likely to discover books via
recommendations regardless of author. Low income individuals are also likely
to discover books through recommendations, but the extent to which
this is the case is more variable by author; for instance, low income individuals who
have read Christie are more likely to discover new books via
recommendations than those who have read Collins.  Among individuals who have read Meyer, those with high incomes
are very likely to discover books at the library, whereas those with low incomes are not.  
Low income individuals appear more likely to discover books via the Internet than high income individuals 
for readers of all authors except Kleypas. In fact, low and high income individuals who have
read Kleypas do not appear to differ in terms of discovery. 

\begin{figure}
\centering

\includegraphics[width=6.5in,height=3in]{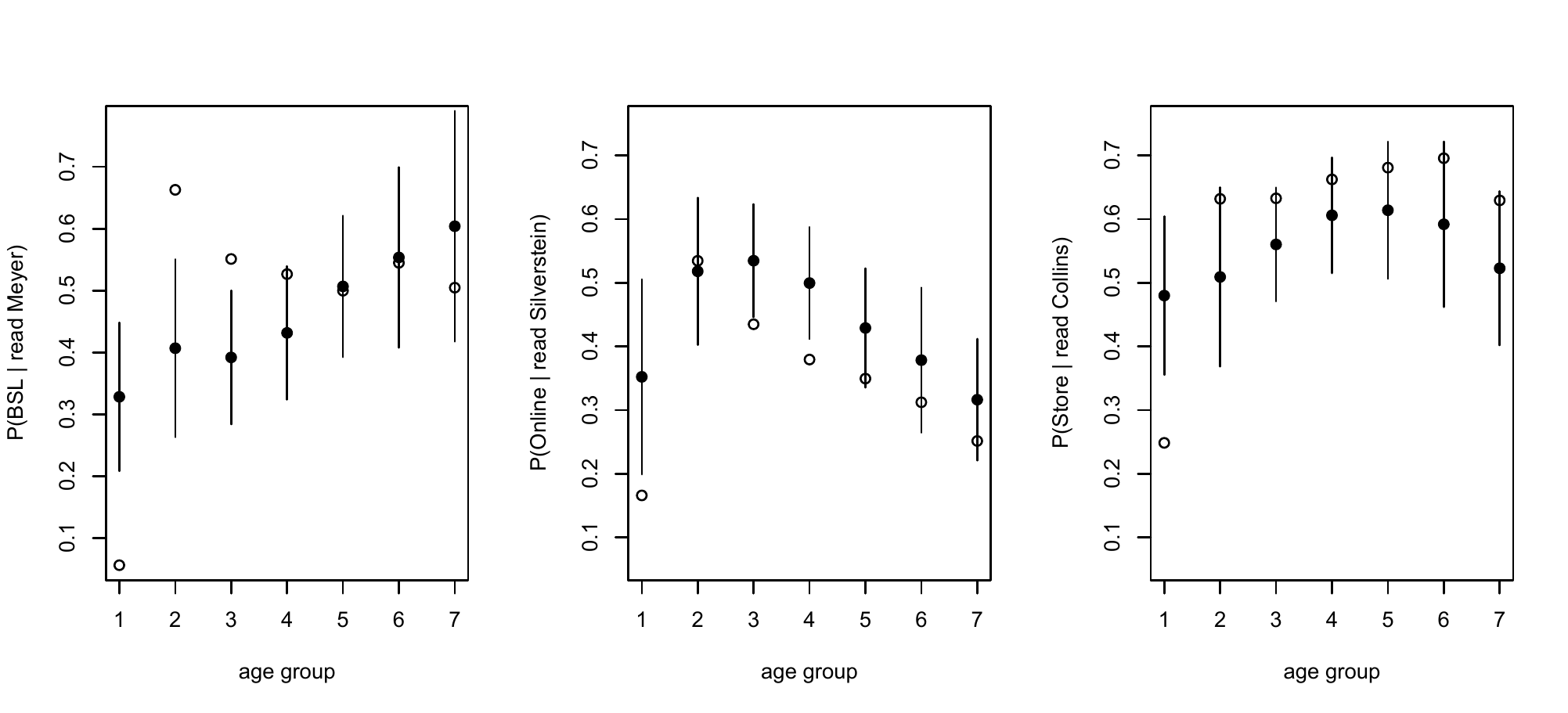}

\caption{Multiple imputation point estimates and $95\%$ confidence intervals for $P(B=\mathrm{yes}|B'=\mathrm{read},\mathrm{age})$ across age groups for three different combinations of mediums $B$ and authors $B'$. Open circles refer to the estimates under the DPMPM model applied without any glue. Left: Probability of discovery via Best Seller List given one has read Meyers. Middle: Probability of discovery Online given one has read Silverstein. Right: Probability of discovery via Bookstores given one has read Collins. }
\label{figure:disc_cond_read_age}
\end{figure}

We also look at author discovery conditional on reading interest and age, as opposed to income. Figure \ref{figure:disc_cond_read_age} displays inference for $P(B=\mathrm{yes}|B'=\mathrm{read},\mathrm{age})$ across age groups for three different combinations of discovery mediums $B$ and authors $B'$. There appears to be an increasing trend in discovery via Best Seller List for those who have read Meyer. In other words, older individuals who have read Meyer are more likely to discover new books through the Best Seller List than younger individuals. Quadratic trends are present for discovery via the Internet for those who have read Silverstein and in discovery via Bookstores for those who have read Collins. As evidence of the impact of glue, 
Figure \ref{figure:disc_cond_read_age} also displays the multiple imputation point estimates obtained from the DPMPM model fit without using the CivicScience data.  In some cases these estimates agree in terms of the trends they suggest (e.g., the middle figure) but sometimes there are fairly stark differences, such as in the leftmost figure. 

\begin{figure}
\centering
\includegraphics[width=6in,height=3.5in]{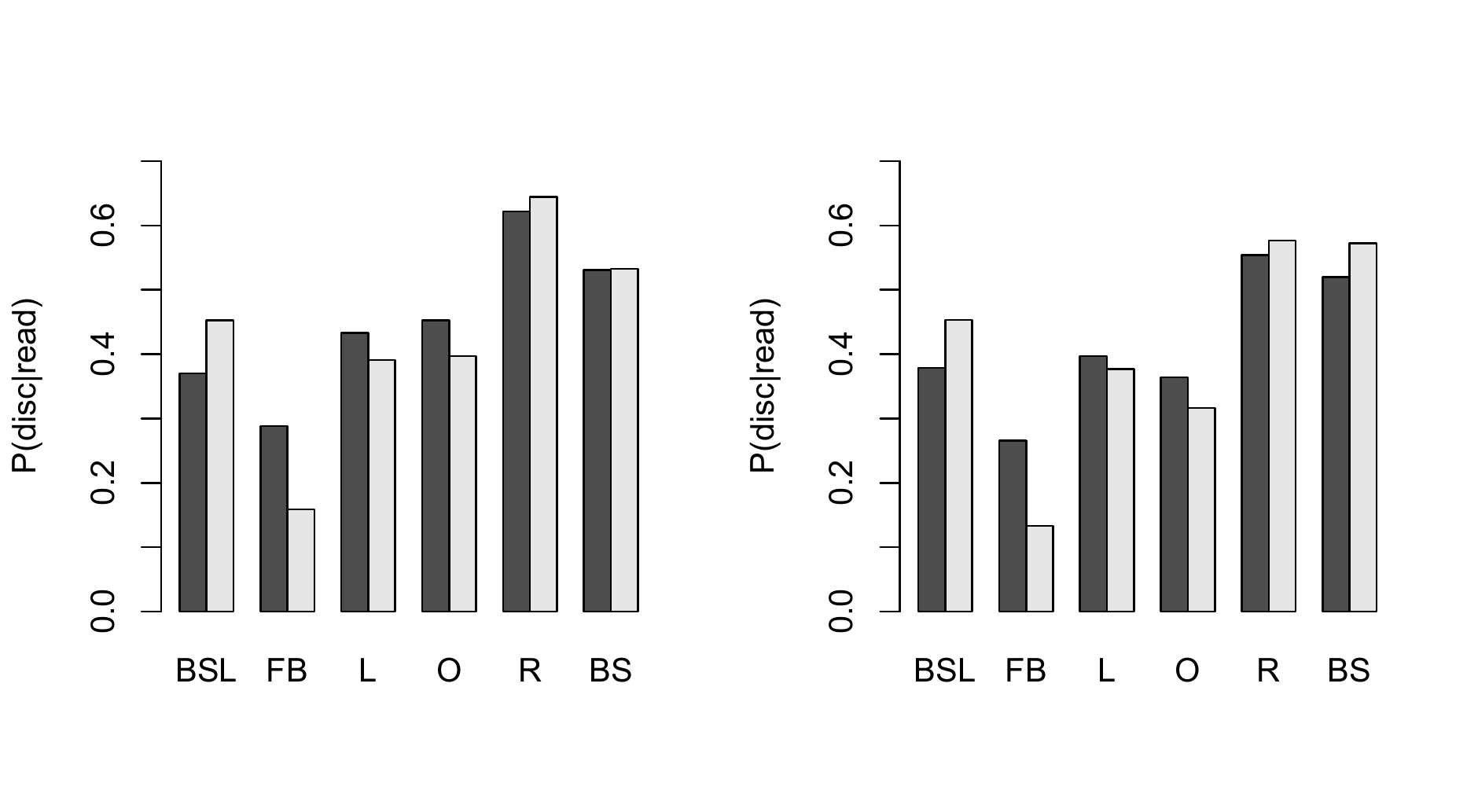}
\caption{Posterior mean estimates for $P(B=\mathrm{yes}\mid B'=\mathrm{read})$ for $B$ representing each of 5 mediums and $B'$ representing Silverstein (dark gray) and Christie (light gray) under the model applied with glue (left) and without glue (right).}
\label{figure:B_cond_Bprime}
\end{figure}

Finally, we estimate the conditional distributions  
$P(B\mid B')$ for particular discovery
mediums and authors.  Figure
\ref{figure:B_cond_Bprime} displays these probability distributions for
authors Silverstein and Christie, under models applied with and without
glue.  
It appears that fans of Silverstein's books use Facebook to find out about new books more
frequently than fans of Christie's books; however, both readerships rely on the Best Seller List equally. 
 We note that the glue impacts inference for even these marginal probabilities.

\section{Concluding remarks}\label{conclusion}

While useful for marketing purposes in their own right, the results of the HarperCollins and CivicScience data fusion 
offer some general lessons about integrating online and traditional survey data.  First, it is possible to improve inferences by collecting glue, 
even when the additional data include only portions of the full joint distribution of interest.  However, crucially, the glue and survey data 
should represent the same distribution.  Second, data from online polling companies like CivicScience, not surprisingly, are likely to be 
not representative on some dimensions.  However, when one believes that conditional distributions in the polling data are reliable, one can 
construct appropriate glue from the conditional distributions, as we did in the HarperCollins data fusion.  Third, it is important to 
understand the limitations of the online data. For example, the CivicScience data include very few young people. Thus, the estimate of 
$P(B\mid A,B')$ from the CivicScience data when $A$ refers to a young person has high variance, so that the glue may not offer adequate information 
about young people.

The simulations with the HarperCollins data also point to interesting directions for future research.  In those 
simulations, adding gender to glue already containing age does not noticeably improve the inferences.  In practice, one would 
expect the cost of collecting glue to increase with the number of variables; hence, in this simulated fusion context, it may not be cost effective to collect gender
as part of the glue.  This suggests a benefit for research on methods for selecting the variables that most improve the accuracy of data fusion, 
taking into account the cost of obtaining those variables.

\newpage

\bibliographystyle{apalike}
\renewcommand\refname{{REFERENCES}}
\bibliography{bibliography}

\newpage
\appendix
\section{Posterior computation}
In order to obtain inference under the hierarchical model, we use a Gibbs sampler to simulate from the posterior distribution $p(\{\phi^{(j)}\},Z,V,\alpha, Y^{(mis)}\mid \mathrm{data})$, where $Y^{(mis)}$ refers to all missing values in $Y_i=(A_i,B_i,B'_i)$ from $D_1$ and $D_2$, and data refers to all observations of $(A_i,B_i,B'_i)$ in $D_1$, $D_2$, and $D_s$. For computational expediency, we need not impute missing values
for $D_s$, as we are simply using this data to inform nonidentifiable
relationships. However, it would be straightforward to impute these missing values just like we impute missing values in $D_1$ and $D_2$. We now describe the posterior full conditionals for all model parameters. 
\subsection*{Full conditional for $Z$}
The mixture allocation variables $Z_i$, for $i=1,\dots,n$, are updated from categorical distributions with probabilities given by
\begin{equation}
p(Z_i=h\mid Y_i,\pi,\phi)=\frac{\pi_h\prod_{j=1}^p\phi^{(j)}_{hY_{ij}}}{\sum_{k=1}^N\pi_k\prod_{j=1}^p\phi^{(j)}_{kY_{ij}}}
\end{equation}
for $h=1,\dots,N$.
For the glue cases, let $J_i$ represent the variables in $\{1,\dots,p\}$ that are observed for glue case $i$. The variable $Z_i$, $i=1,\dots,n_s$, is updated from a categorical distribution with
\begin{equation}
p(Z_i=h\mid Y_i,\pi,\phi)=\frac{\pi_h\prod_{j\in J_i}\phi^{(j)}_{hY_{ij}}}{\sum_{k=1}^N\pi_k\prod_{j\in J_i}\phi^{(j)}_{kY_{ij}}}
\end{equation}
for $h=1,\dots,N$.

\subsection*{Full conditional for $\{\phi^{(j)}\}$}
To update $\phi_{h}^{(j)}$, for $h=1,\dots,N$, and $j=1,\dots,p$, sample from a Dirichlet distribution:
\begin{equation}
p(\phi_h^{(j)}\mid Y^{(obs)},Z)\propto \mathrm{Dirichlet}\left(\phi_h^{(j)}; 1+\sum_{\{i:Z_i=h\}}1(Y_{ij}=1),\dots,1+\sum_{\{i:Z_i=h\}}1(Y_{ij}=d_j)\right),
\end{equation}
where the summations are over all survey and glue cases, $i\in\{1,\dots,n+n_s\}$.

\subsection*{Full conditional for $V$}
The stick-breaking proportions $V_h$, for $h=1,\dots,N-1$, can be sampled from Beta distributions:
\begin{equation}
p(V_h\mid\alpha,Z)\propto \mathrm{Beta}\left(V_h; M_h+1,\alpha+\sum_{j=h+1}^NM_j\right),
\end{equation}
where $M_h=\sum_{i=1}^{n+n_s}1(Z_i=h)$. Fixing $V_N=1$, the probabilities $\pi$ are given by $\pi_1=V_1$ and $\pi_h=V_h\prod_{j=1}^{h-1}(1-V_j)$ for $h=1,\dots,N$.

\subsection*{Full conditional for $\alpha$}
The DP precision parameter $\alpha$ can be sampled from a Gamma distribution:
\begin{equation}
p(\alpha\mid V)\propto \mathrm{Gamma}\left(\alpha;N+a_{\alpha}-1,b_{\alpha}-log(\pi_N)\right).
\end{equation}

\subsection*{Imputing $Y^{(mis)}$}
Missing $Y_{ij}$ in $D_1$ and $D_2$ can be imputed by sampling from categorical distributions with the form given in equation (\ref{model:Y}).
\end{document}